\documentclass[12pt,preprint]{aastex}

\shorttitle{Generalized Tests for GRBs}
\shortauthors{Schaefer \& Collazzi}

\begin{document}

\title{Generalized Tests for Eight GRB Luminosity Relations}

\author{Bradley E. Schaefer}
\author{Andrew C. Collazzi}
\affil{Physics and Astronomy, Louisiana State University,
    Baton Rouge, LA, 70803}

\begin{abstract}

Long duration Gamma-Ray Bursts (GRBs) have eight luminosity relations where observable burst properties can yield the burst luminosity and hence distance.  This turns GRBs into useful tools of cosmology.  Recently, two tests have been proposed (by Nakar \& Piran and by Li) for which one of the eight relations is claimed to have significant problems.  In this paper, we generalize these tests and apply them to all eight GRB luminosity relations.  (a) All eight relations pass the Nakar \& Piran test after accounting for the uncertainties on the data and the dispersions of the correlations.  (b) All eight relations are good when the GRB redshifts are {\it known}, for example for calibration of the relations and for GRB Hubble diagram purposes.  (c) We confirm the earlier results that the $E_{\gamma ,iso} - E_{peak}$ Amati relation must produce very large error bars whenever an {\it unknown} redshift being sought is $\gtrsim 1.4$.  (d) The $E_{\gamma} - E_{peak}$ relation of Ghirlanda et al. must produce very large error bars whenever an {\it unknown} redshift being sought is $\gtrsim 3.4$.  (e) The other six relations have no problem at all from the ambiguity test of Li. 

\end{abstract}
\keywords{Gamma-Ray: Bursts}



\section{Introduction}

	Long-duration Gamma-Ray Bursts (GRBs) have eight separate luminosity relations, where a measured photometric or spectroscopic property is correlated with the burst's luminosity (Norris, Marani, \& Bonnell 2000; Fenimore \& Ramirez-Ruiz 2000; Schaefer 2003; Amati et al. 2002; Ghirlanda, Ghisellini, \& Lazzati 2004; Schaefer 2002; Firmani et al. 2006).  Physical explanations for these relations have been proposed (Schaefer 2004; M\'{e}sz\'{a}ros et al. 2002; Kobayashi, Ryde, \& MacFadyen 2002; Schaefer 2003; Rees \& M\'{e}sz\'{a}ros 2005; Schaefer 2002; Thompson, M\'{e}sz\'{a}ros, \& Rees 2006; Eichler \& Levinson 2004; Liang, Dai, \& Wu 2004; Giannios \& Spruit 2006).  The relations return luminosities with varying degrees of accuracy.  As the returned luminosities from each of the eight relations are largely independent, they are best combined to produce a single weighted-average luminosity with a surprisingly good accuracy.  This combined luminosity along with the burst's brightness can give a luminosity distance to the GRB, and then paired with a fiducial cosmology to yield a redshift of the burst.  With our sample of 69 GRBs with spectroscopic redshifts, we have already demonstrated that the accuracy is 26\% in redshift (Schaefer 2006).  The scatter of our derived redshifts about the spectroscopic redshifts is a good Gaussian distribution with a reduced chi-square of nearly unity, so this demonstrates that our derived error bars are realistic.  With this, we have a method to derive distances to all GRBs with good confidence to within the quoted error bars.
	
	Nakar \& Piran (2005) have proposed a new test of one particular GRB luminosity relation.  The tested relation is the one between the GRB isotropic energy in the gamma-ray band, $E_{\gamma ,iso}$, and the observed photon energy of the peak in the $\nu F_{\nu}$ spectrum, $E_{peak}$, as first found by Amati et al. (2002).  Their idea was to set up an equality between the energy based on the observed fluence and the energy derived from the luminosity relation, move all the redshift dependent terms to one side and the observables to the other side of the equation.  From here, a maximum can be found on the redshift side and compared with the quantity calculated from the observables for many GRBs.  The Nakar \& Piran test is then to see whether the derived quantities from the observables exceeds this maximum possible value (at least to within the error bars).  When applied to a sample of BATSE bursts, they found that $\sim$25\% of their samples violated this simple requirement.  Subsequently, Band \& Preece (2005) found that 88\% of their sample of BATSE bursts violated the Nakar \& Piran test.  This would be a serious blow against the validity of one of the luminosity relations.  Amati (2006) has strongly defended this particular luminosity relation, mainly on the grounds that the luminosity relation is highly significant, broadly applicable, and fits a well-defined calibration.  Ghirlanda, Ghisellini, \& Firmani (2005) have explained the discrepancy as simply that Nakar \& Piran used an old version of the Amati relation (based on just nine bursts) and did not allow for the real uncertainties.
	
	Li (2006) has applied a similar analysis to the $E_{\gamma ,iso} - E_{peak}$ Amati relation to demonstrate that the returned redshifts for any burst is actually ambiguous with two possible values.  Li finds the maximum to occur at a redshift of 3.8, such that, for example, a $z=2.8$ burst has identical properties as a $z=5$ burst.  In addition, around the redshift of the maximum, the derived error bar should be very large. For realistic error bars, Li finds that the Amati relation will return redshifts with very large uncertainties for bursts with $z \gtrsim 1.4$.
	
	With this, we have what can be perceived as a significant challenge to the reliability, ambiguity, and utility of one of the eight luminosity relations.  The obvious task is to generalize this test to all eight of the luminosity relations.  Indeed, Band \& Preece have already applied the Nakar \& Piran test to the $E_{\gamma} - E_{peak}$ relation of Ghirlanda, Ghisellini, \& Lazzati (2004), and found only a small fraction of violators (1.6\%).  In this paper, we generalize the tests of Nakar \& Piran and Li to all eight GRB luminosity relations.  The importance of these tests is that they bear on the validity and accuracy of all the relations, and the utility of these relations is a prerequisite for getting cosmology from GRBs.
	
\section{The Generalized Test}

	We will now present the same derivation as used in Nakar \& Piran (2005) and Li (2006), but with a generalized notation that can handle all eight luminosity relations.
	
	The luminosity relations connect  a measure of the burst's luminosity (notated as $\cal L$) with a luminosity indicator (notated as $\cal I$).  The luminosity, $\cal L$, can be the isotropic luminosity, $L$, the isotropic energy emitted in gamma rays, $E_{\gamma,iso}$, or the $E_{\gamma,iso}$ value corrected by the beaming factor ($F_{beaming}$) from the jet, $E_{\gamma}$.  The luminosity indicator, $\cal I$, can be the spectral lag $\tau_{lag}$ (Norris, Marani, \& Bonnell 2000), the variability $V$ (Fenimore \& Ramires-Ruiz 2000), the $E_{peak}$ from the spectrum, the minimum rise time in the light curve $\tau_{RT,min}$, or the number of peaks in the light curve $N_{peak}$.  In addition, Firmani et al. (2006) take a particular combination involving $E_{peak}$ and $T_{0.45}$ (the duration over which the brightest portion of the light curve emits 45\% of the fluence) to be a luminosity indicator, as given by $E_{peak}^{1.62} T_{0.45}^{-0.49}$.  All of these indicators have to be corrected from their observed values to the values in the GRB rest frame by multiplication by $1+z$ raised to a power $Q$.  The relations are all simple power laws with indices $m$ and constants $\cal A$.  The luminosity relations can be expressed as 
\begin{equation}
{\cal L} = {\cal A} [ {\cal I} (1+z)^Q]^m.
\end{equation}
The eight relations have their particular values for $\cal L$, $\cal A$, $\cal I$, $Q$, and $m$ given in Table 1.

	The $\cal L$ for each burst of known redshift can be determined from the observed brightness, $\cal B$.  The $\cal B$ value will either be the bolometric peak flux, $P_{bolo}$, the bolometric fluence, $S_{bolo}$, or the beaming corrected fluence, $S_{bolo}F_{beaming}$, depending on the value for $\cal L$.  The inverse square law of light can then be expressed as 
\begin{equation}
{\cal L} = 4 \pi d_L^2 {\cal B} (1+z)^{-B}.
\end{equation}
The luminosity distance, $d_L$, is related to the redshift, $z$, through the usual integral; for which we have throughout this paper assumed a flat universe with $\Omega_M=0.27$ and an unchanging cosmological constant of $w=-1$.  When we are dealing with fluences and burst energies, a factor of $(1+z)^{-1}$ is need to correct for time dilation.  The corresponding values for $\cal B$ and $B$ are presented in Table 1 for each luminosity relation.

	We can use equations 1 and 2 to eliminate $\cal L$.  With this, we can place all the redshift-dependent terms onto the left side of the equation and all the observable quantities on the right side of the equation.  Finally, we can multiply both sides by $(H_0/c)^2$ so as to make both sides dimensionless and of reasonable magnitude.  Thus,
\begin{equation}
(H_0/c)^2 d_L^2 (1+z)^{-Qm-B} = [(H_0/c)^2 /4\pi] ({\cal A ~ I}^m / {\cal B}).
\end{equation}
We will be making frequent reference to both the left and right sides of this equation separately, so we will notate the two sides separately.  This will be
\begin{equation}
{\cal F}({\cal I}, {\cal B}) = [(H_0/c)^2 /4\pi] ({\cal A ~ I}^m / {\cal B}),
\end{equation}
\begin{equation}
{\cal F}(z) = (H_0/c)^2 d_L^2 (1+z)^{-Qm-B}.
\end{equation}
With this, we expect ${\cal F}({\cal I}, {\cal B}) = {\cal F}(z)$.

	It is convenient to define a maximum value for ${\cal F}(z)$, which we will label as ${\cal F}_{max}$.  For two of the relations, the ${\cal F}(z)$ comes to a simple maximum at some moderate redshift value, $z_{peak}$ (see Table 1).  For the other relations, the ${\cal F}(z)$ keeps rising as the redshift increases out past where any GRB might lie.  In the spirit of this test, we know that these ${\cal F}(z)$ values for all observed bursts must be less that the value at the maximum GRB redshift.  From Bromm \& Loeb (2002), we know that GRBs cannot be made at $z>20$ or so.  Thus, for the six relations with no $z_{peak}$, we take ${\cal F}_{max}$ to be the value of ${\cal F}(z=20)$.  The values of ${\cal F}_{max}$ are given in Table 1.  So in the most general terms, the Nakar \& Piran test is whether ${\cal F}({\cal I}, {\cal B})/{\cal F}_{max} > 1$ for observed GRBs, while the Li test is whether $z_{peak} \lesssim 10$.
	
	To extend the Nakar \& Piran test, we have chosen to use a set of 69 GRBs with spectroscopic redshifts for which all the required data have already been extracted and reduced for purposes of making a GRB Hubble diagram.  Complete details on the selection of these bursts, their redshifts, and all their observed properties are presented in Schaefer (2006).  These bursts were the ones used to derive the best fit luminosity relations (Schaefer 2006) as expressed by Equation 2 and the parameters in Table 1.  The scatter of the observed ${\cal L}$ about the value derived from the observed ${\cal I}$ varies widely amongst the relations, with the scatter generally being larger than the error bars from measurement errors alone.  Schaefer (2006) has shown that this systematic scatter is not apparently dependent on redshift or luminosity, and its magnitude is consistent with arising from the expected scatter due to Malmquist and gravitational lensing effects.  To account for this systematic error, we add in quadrature the values listed in Table 1.  The result is that the combined redshifts (from all available luminosity indicators) have a one-sigma scatter of 26\% when compared to the spectroscopic redshifts (see Figure 9 of Schaefer 2006).  In addition, the reduced chi-square of this comparison is close to unity, demonstrating that the derived error bars are reliable.  It is this analysis which validates the use of the luminosity relations to get burst redshifts which are reliable to within the quoted error bars (typically 26\%).

\section{Test for Ambiguity in Deriving Redshifts}
		
	The utility of the luminosity relations is that we can go from observed quantities to the distance.  In more detail, we can calibrate the relations with GRBs of known redshift so as to derive $\cal A$ and $m$, measure $\cal I$ for each burst, calculate ${\cal F}({\cal I}, {\cal B})$, set ${\cal F}(z)$ as being equal to ${\cal F}({\cal I}, {\cal B})$, then determine the luminosity distance and redshift that produces the ${\cal F}(z)$.  The problem that Li (2006) pointed out is that this procedure is ambiguous for the $E_{\gamma ,iso} - E_{peak}$ Amati relation, because there are always {\it two} distances/redshifts that produce the observed value ${\cal F}({\cal I}, {\cal B})$.  That is, there will always be a redshift below $z_{peak}$ and a redshift above $z_{peak}$ for which ${\cal F}({\cal I}, {\cal B}) = {\cal F}(z)$.  To take a specific example, for the $E_{\gamma ,iso} - E_{peak}$ Amati relation, a $z=1.5$ burst could be confused with a $z=10.6$ burst, and a $z=1$ burst could be confused for a $z=18.5$ burst; at least in principle.  More importantly, Li points out that the error bars on the derived redshift will be quite large when $z \sim z_{peak}$, with the reason being that ${\cal F}(z)$ changes little with redshift.  Again with the $E_{\gamma ,iso} - E_{peak}$ Amati relation, ${\cal F}(z)$ is within 10\% of ${\cal F}_{max}$ for $2.1<z<7.0$.  Given a one-sigma uncertainty in $\log {\cal L}$ of $\sim 0.15$ (Amati 2006), any GRB with $z>1.4$ will be within one-sigma of $z_{peak}$.  As such, the test of Li puts severe limits on the utility of one of the luminosity relations for purposes of deriving redshifts.
	
	Amati (2006) defends his relation by pointing to the tight calibration curves over a large dynamic range of luminosities.  This is indeed a strong defense if the question is about the existence and the fit parameters (i.e., ${\cal A}$ and $m$) for the $E_{\gamma ,iso} - E_{peak}$ relation.  This is because bursts with {\it known} redshifts will have a unique and well determined value for ${\cal F}(z)$.  But this does not work in the other direction, as a known value of ${\cal F}(z)$ (from the observed ${\cal F}({\cal I}, {\cal B})$) does not produce a unique or necessarily well-determined value for the redshift.  So both sides are right; the $E_{\gamma ,iso} - E_{peak}$ relation certainly exists, while the relation fails in practice for determining the redshift if $z \gtrsim 0.9$.
	
	We now extend the test of Li to all eight luminosity relations.  This can be done by calculating the values of ${\cal F}(z)/{\cal F}_{max}$ from $0<z<20$, as we have plotted in Figure 1.  The curve that rises the fastest is for the $E_{\gamma ,iso} - E_{peak}$ Amati relation, and we immediately see why it runs into trouble with the test of Li.  The reason is that $z_{peak}$ is in the range of redshift where many GRBs are seen, so that a horizontal line corresponding to ${\cal F}({\cal I}, {\cal B})/{\cal F}_{max}$ intersects the curve in two places.  Also, the nearly flat part of the curve (around $z_{peak}$) implies that a given measured value of ${\cal F}({\cal I}, {\cal B})/{\cal F}_{max}$ (with the usual uncertainties) will fit the curve over a large range of redshifts.  With this, we see that a luminosity relation will have trouble with the test of Li only if it has $z_{peak} \lesssim 10$ or if there is a near-flat portion.
	
	Six of the relations have no problem with the test of Li, as they are monotonically rising fast out to high redshifts.  Other than the $E_{\gamma ,iso} - E_{peak}$ relation, only one relation has apparent problems, the $E_{\gamma} - E_{peak}$ relation of Ghirlanda, Ghisellini, \& Lazzati (2004).  In this case, $z_{peak}=7.4$.  So, for example, we might be in danger of confusing a $z=3.2$ burst for a $z=20$ burst.  More importantly, high redshift bursts must have a large uncertainty in any derived redshift.  For a one-sigma scatter in $\log {\cal L}$ of $\sim 0.06$ (Ghirlanda, Ghisellini, \& Lazzati 2004), all bursts with redshift $z \gtrsim 3.4$ will be within one-sigma of each other.
	
	In general, the ambiguities in derived redshifts for the $E_{\gamma ,iso} - E_{peak}$ and $E_{\gamma} - E_{peak}$ relations can always be resolved.  For example, with the $E_{\gamma} - E_{peak}$ relation, the high redshift branch will return $z$ values that can be rejected due to afterglow being seen at optical wavelengths (as required to observe a jet break) which would be shorter than the Lyman limit.  Also, the lower possible redshift will always be much more likely than the higher possible redshift simply due to the rarity of very high luminosity events in the GRB luminosity function.  But the general solution is to have multiple luminosity indicators, for which the various indicators will overlap for only one solution.  In all, even though there is formally an ambiguity for two of the relations, the ambiguities will always be resolved in practice.
	
	One of the uses of the luminosity indicators is in the construction of a GRB Hubble diagram from bursts with known redshifts.  The problems noted by Li are not relevant in this case, as the {\it known} redshifts allow us to derive a unique and well-determined value of ${\cal F}(z)$ (for the given cosmology).
	
	In summary: (a) Li's test is easily passed for six of the relations for all questions, (b) the redshift ambiguity will always be resolved in practice for the other two relations, (c) the $E_{\gamma ,iso} - E_{peak}$ and $E_{\gamma} - E_{peak}$ relations cannot return accurate derived redshifts for $z \gtrsim 1.4$ and $z \gtrsim 3.4$ respectively, (d) all eight luminosity relations have no problems for questions (such as the GRB Hubble diagram) involving bursts of {\it known} redshifts.
	
\section{Test for Violators}

	Nakar \& Piran (2005) and Band \& Preece (2005) report that a large fraction of GRBs violate the idealized requirement that ${\cal F}({\cal I}, {\cal B})/{\cal F}_{max} \leq 1$ for the $E_{\gamma ,iso} - E_{peak}$ relation.  Ghirlanda, Ghisellini, \& Firmani (2005) point out that these violations arise only due to the use of an old calibration of the Amati relation and due to unrealistically small adopted error bars.  For whatever resolution, the test of Nakar \& Piran should be extended to all eight luminosity relations.
	
	For the 69 GRBs, we calculate ${\cal F}({\cal I}, {\cal B})/{\cal F}_{max}$ for each burst.  Then we count how many burst have this quantity greater than unity, and hence are violators in the Nakar \& Piran test.  For each luminosity relation, the fraction of violators are given in the last column of Table 1.  Three of the relations have zero violators.  Three other relations each have only two bursts that are just barely in violation, where the violators are all within 0.6-sigma of the expected value of ${\cal F}(z)/{\cal F}_{max}$.  Thus, these six relations pass the Nakar \& Piran test.  But two relations (the $E_{\gamma ,iso} - E_{peak}$ and $E_{\gamma} - E_{peak}$ relations) have substantial fractions of violators, and it is no coincidence that these are the same relations that have $z_{peak} \lesssim 10$.
	
	For the two relations with the most violators, we have plotted the observed ${\cal F}({\cal I}, {\cal B})/{\cal F}_{max}$ and the theoretical ${\cal F}(z)/{\cal F}_{max}$ as a function of redshift (see Figure 2).  If the luminosity indicators were perfect, then all the observed points would fall along the smooth curve.  The plotted error bars as well as the apparent scatter about the model curve illustrate the typical scatter that arises in each relation.  Violators of the Nakar \& Piran test are those GRBs that are higher than zero on this log-scale.
	
	By using a sample of bursts with spectroscopic redshifts, we can better see what is going on.  In particular, we see that the most of the GRBs have redshifts where $\log [{\cal F}(z)/{\cal F}_{max}] \approx 0$, and so the normal scatter in the relation naturally creates many violators.  At $z_{peak}$, we expect half the bursts to be violators.  And for redshifts of $z>1.4$, we have ${\cal F}(z)/{\cal F}_{max}$ close to unity, so somewhat less than 50\% of the bursts should be violators.  We see from Figure 2 that the violators are simply the upper tail of a normal distribution, and hence do not significantly violate the $E_{\gamma ,iso} - E_{peak}$ relation.
	
	For the $E_{\gamma} - E_{peak}$ relation of Ghirlanda, Ghisellini, \& Lazatti (2004), Figure 2 shows a similar situation with only five violators, all with $<1-\sigma$ deviations.  The curve is near zero for the higher redshifts, the violators are all at high redshift, and the scatter about the curve is normally distributed.  Again, the existence of violators appears to be a simple consequence of the expected scatter from both systematic and observational errors.
	
	In summary of our generalized Nakar \& Piran test, we find that all eight luminosity relations satisfy the test.  In particular, while some bursts have ${\cal F}({\cal I}, {\cal B})/{\cal F}_{max} > 1$, this is an expected consequence of ordinary scatter about the best fit relation.

\clearpage

\begin{deluxetable}{lllllllllll}
\tabletypesize{\scriptsize}
\tablecaption{Luminosity Relations 
\label{tbl1}}
\tablewidth{0pt}
\tablehead{
\colhead{Relation}   &
\colhead{$\cal I$}   &
\colhead{m}  &
\colhead{Q}  &
\colhead{$\cal A$}  &
\colhead{$\cal B$}  &
\colhead{B}  &
\colhead{$\cal F$(z)$_{max}$}  &
\colhead{$z_{peak}$}  &
\colhead{$\sigma_{\log {\cal L}}$}  &
\colhead{Violators\tablenotemark{a}}  
}
\startdata
$\tau_{lag}$-$L$	&	$\tau_{lag}$	&	-1.01	&	-1	&	$1.8 \times 10^{51}$	&	$P_{bolo}$	&	0	&	139	&	$\gg 20$	&	0.39	&	0\%  \\
$V$-$L$	&	$V$	&	1.77	&	1	&	$3.2 \times 10^{55}$	&	$P_{bolo}$	&	0	&	13.7	&	$\gg 20$	&	0.40	&	4\%\tablenotemark{b}  \\
$E_{peak}$-$L$	&	$E_{peak}$	&	1.68	&	1	&	$1.1 \times 10^{48}$	&	$P_{bolo}$	&	0	&	18.0	&	$\gg 20$	&	0.36	&	3\%\tablenotemark{c}  \\
$E_{peak}$-$E_{\gamma,iso}$	&	$E_{peak}$	&	2.04	&	1	&	$8.5 \times 10^{47}$	&	$S_{bolo}$	&	1	&	0.56	&	3.8	&	0.15	&	44\%\tablenotemark{d}  \\
$E_{peak}$-$E_{\gamma}$	&	$E_{peak}$	&	1.61	&	1	&	$4.5 \times 10^{46}$	&	$S_{bolo}F_{beaming}$	&	1	&	1.23	&	7.4	&	0.16	&	19\%\tablenotemark{e}  \\
$\tau_{RT}$-$L$	&	$\tau_{RT}$	&	-1.21	&	-1	&	$2.1 \times 10^{51}$	&	$P_{bolo}$	&	0	&	75.4	&	$\gg 20$	&	0.47	&	0\%  \\
$N_{peak}$-$L$	&	$N_{peak}$	&	2	&	0	&	$2.1 \times 10^{50}$	&	$P_{bolo}$	&	0	&	3002	&	$\gg 20$	&	0	&	0\%  \\
$E_{peak}T_{0.45}$-$L$	&	$E_{peak}^{1.62}T_{0.45}^{-0.49}$	&	1	&	1.91	&	$3.2 \times 10^{48}$	&	$P_{bolo}$	&	0	&	8.95	&	$\gg 20$	&	0.16	&	4\%\tablenotemark{f}  \\
\enddata
    
\tablenotetext{a}{$\log [{\cal F}({\cal I}, {\cal B})/{\cal F}_{max}]$ values for the violators are given with footnotes.}  
\tablenotetext{b}{$0.46 \pm 1.01$ for GRB 050502 and $0.19 \pm 1.17$ for GRB 060526.}
\tablenotetext{c}{$0.48 \pm 0.93$ for GRB 050904 and $0.10 \pm 1.11$ for GRB 060605.}
\tablenotetext{d}{$1.22 \pm 0.45$ for GRB 970508, $0.49 \pm 0.42$ for GRB 971214, $0.57 \pm 1.04$ for GRB 980613, $0.29 \pm 0.41$ for GRB 980703, $0.02 \pm 0.42$ for GRB 990123, $0.13 \pm 0.37$ for GRB 000210, $0.62 \pm 0.41$ for GRB 000911, $0.04 \pm 0.62$ for GRB 020405, $0.16 \pm 0.76$ for GRB 021004, $0.24 \pm 0.70$ for GRB 030115, $0.46 \pm 1.19$ for GRB 050406, $0.37 \pm 0.89$ for GRB 050502, $0.55 \pm 0.47$ for GRB 050603, $1.16 \pm 0.51$ for GRB 050820, $0.92 \pm 0.57$ for GRB 050904, $0.06 \pm 0.53$ for GRB 050908, $0.80 \pm 0.49$ for GRB 050922, $0.51 \pm 0.64$ for GRB 051109, $0.61 \pm 0.67$ for GRB 060108, $0.43 \pm 0.81$ for GRB 060116, $0.16 \pm 0.57$ for GRB 060124, $0.40 \pm 0.55$ for GRB 060206, $0.43 \pm 0.96$ for GRB 060223, $0.25 \pm 0.43$ for GRB 060418, $0.52 \pm 0.77$ for GRB 060502, $0.61 \pm 0.54$ for GRB 060604, $1.27 \pm 0.69$ for GRB 060605, and $0.27 \pm 0.57$ for GRB 060607.}
\tablenotetext{e}{$0.07 \pm 0.62$ for GRB 050318, $0.26 \pm 0.88$ for GRB 050505, $0.18 \pm 0.58$ for GRB 050904, $0.20 \pm 0.55$ for GRB 060124, and $0.70 \pm 0.74$ for GRB 060210.}
\tablenotetext{f}{$0.19 \pm 0.55$ for GRB 050904 and $0.30 \pm 0.82$ for GRB 060605.}
    
\end{deluxetable}

\clearpage

\begin{figure}
\epsscale{.80}
\plotone{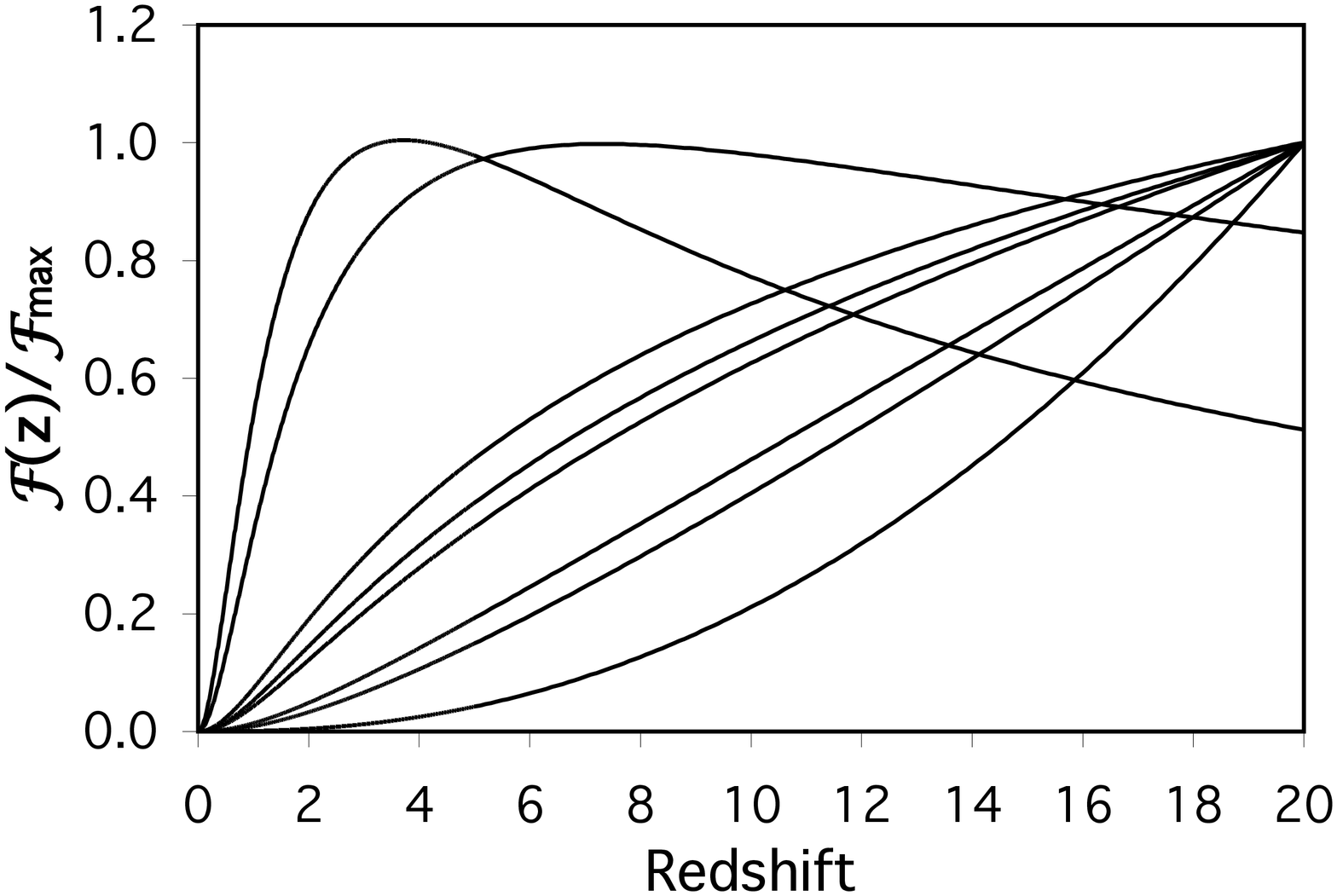}
\caption{
${\cal F}(z)/{\cal F}_{max}$ for the eight luminosity relations.  The eight curves, from left to right where they intersect the horizontal line at 0.4 are for the $E_{peak}$-$E_{\gamma,iso}$, $E_{peak}$-$E_{\gamma}$, $E_{peak}T_{0.45}$-$L$, $V$-$L$, $E_{peak}$-$L$, $\tau_{RT}$-$L$, $\tau_{lag}$-$L$, and $N_{peak}$-$L$ relations.  The last six of these relations all meet at 1.0 for $z=20$ as the ${\cal F}_{max}$ value was taken at $z=20$ for these relations with $z_{peak} \gg 20$.  When dealing with GRBs of {\it unknown} redshift, the value of ${\cal F}({\cal I}, {\cal B})/{\cal F}_{max}$ is derived from the data, the ${\cal F}(z)/{\cal F}_{max}$ is set equal to this, and then the plot is used to determine the redshift of the GRB.  Li (2006) realized that the $E_{peak}$-$E_{\gamma,iso}$ relation is ambiguous (in that two redshifts will both fit the observations) and that the uncertainty in the derived redshift will be large for bursts near the peak in the curve.  With our generalized test, we see that the $E_{peak}$-$E_{\gamma}$ relation also has the same problem, but at higher redshifts.  Given the typical uncertainties, this means that the $E_{peak}$-$E_{\gamma,iso}$ and $E_{peak}$-$E_{\gamma}$ relations cannot be used with any accuracy to determine the redshifts of GRBs with $z \gtrsim 1.4$ and $z \gtrsim 3.4$ respectively.  These problems arise due to the ${\cal F}(z)/{\cal F}_{max}$ function having a maximum (at $z_{peak}$) at redshifts below $\sim 10$.  However, all other luminosity relations easily pass the test of Li.  Also, when the GRB redshift is known from optical spectroscopy, the ${\cal F}(z)/{\cal F}_{max}$ value will be uniquely and accurately determined, so all eight luminosity relations are fine for questions like  the calibration of the relations and the construction of the GRB Hubble diagram.
}

\end{figure}

\clearpage

\begin{figure}
\epsscale{1.10}
\plottwo{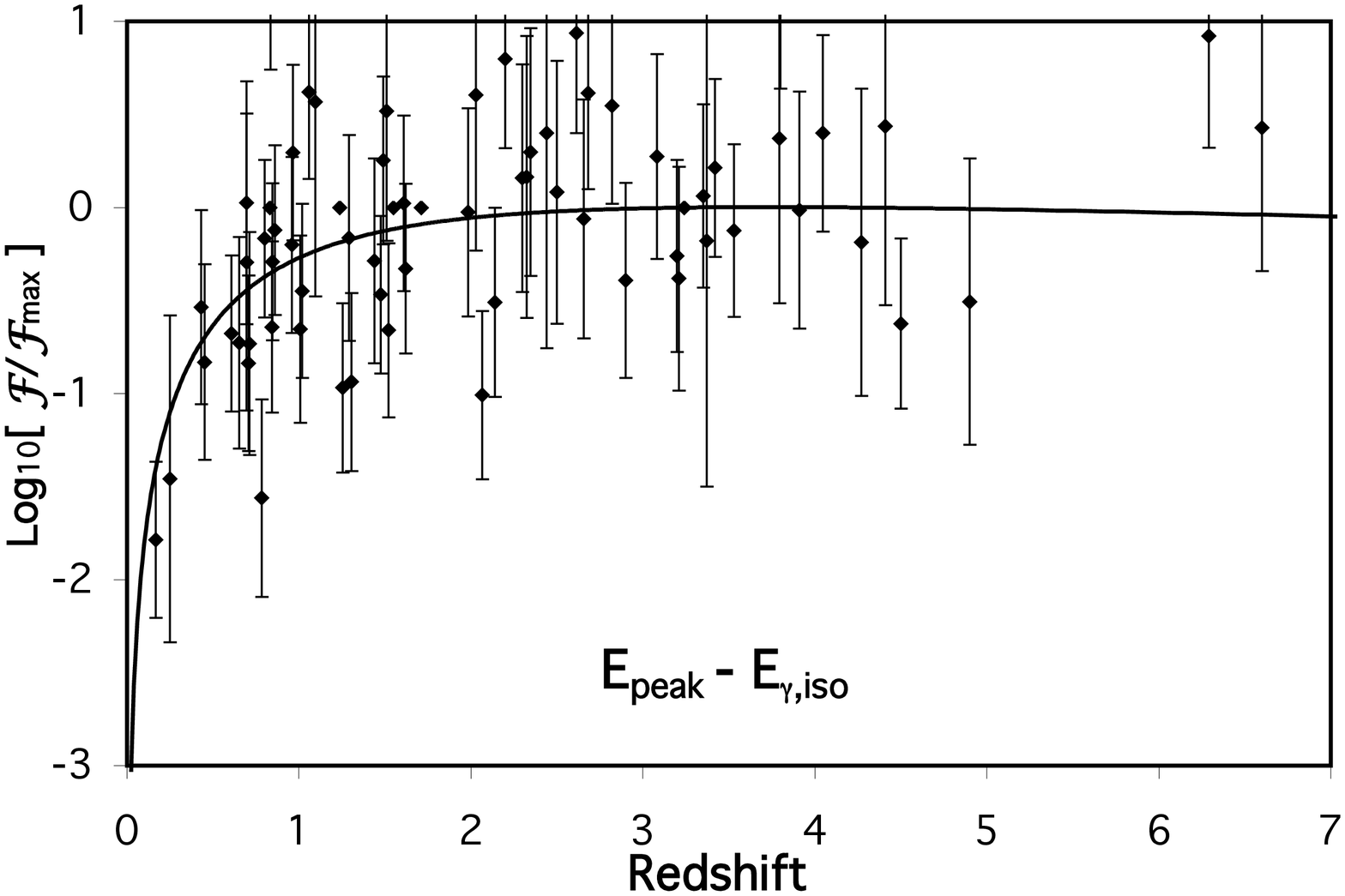}{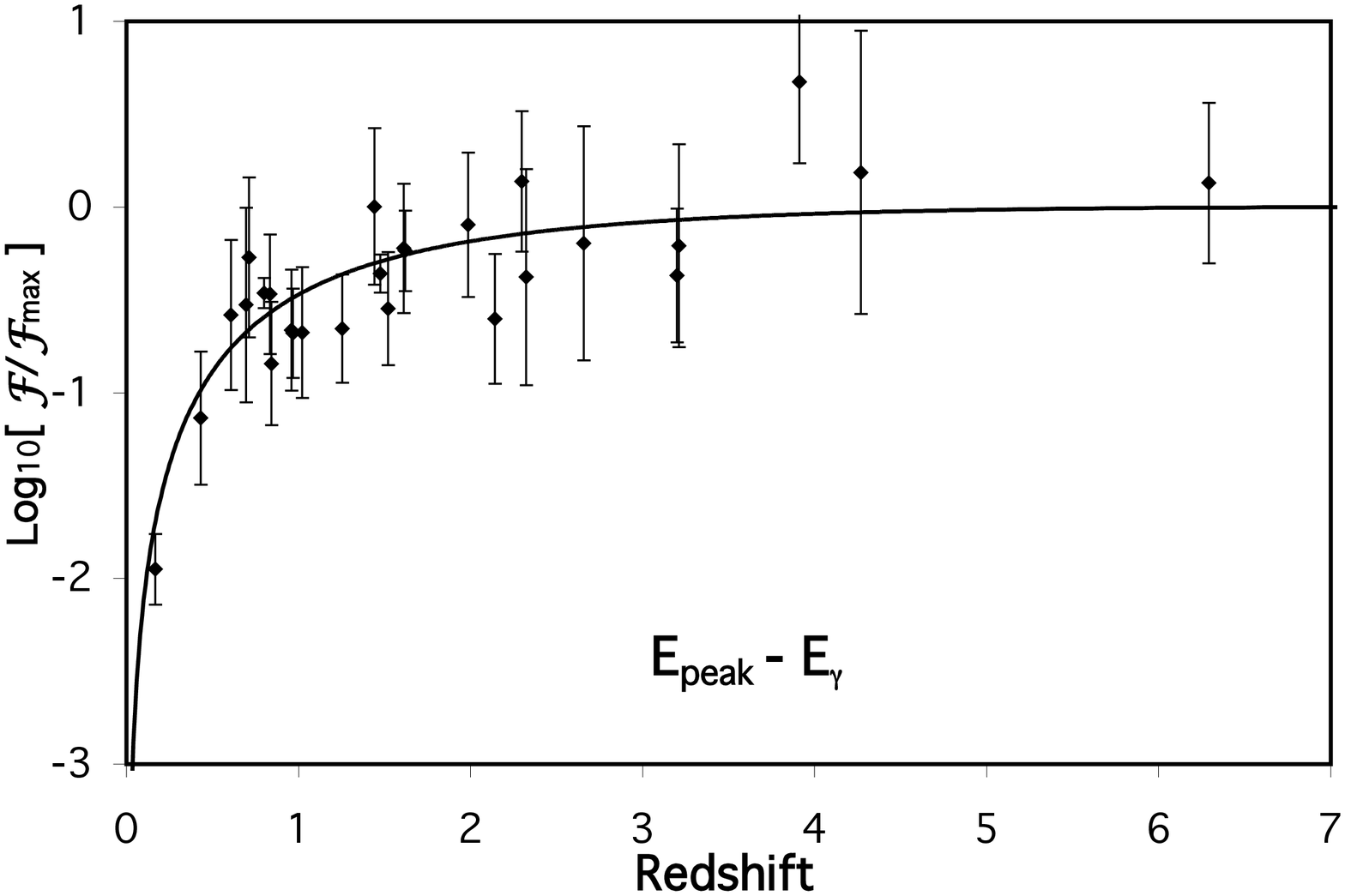}
\caption{
${\cal F}/{\cal F}_{max}$ for two relations.  The observed ${\cal F}({\cal I}, {\cal B})/{\cal F}_{max}$ values for each burst are over-plotted on the theoretical ${\cal F}(z)/{\cal F}_{max}$ curve for the $E_{peak}$-$E_{\gamma,iso}$ Amati relation (left panel) and the $E_{peak}$-$E_{\gamma}$ relation of Ghirlanda et al. (right panel).  Bursts which violate the Nakar \& Piran test are those with $\log [{\cal F}({\cal I}, {\cal B})/{\cal F}_{max}] > 0$ on this plot.  For these two relations only, the theoretical curve is close to the limit over much of the redshift range of observed bursts.  With the normal scatter apparent in these plots, we expect roughly half of the GRBs to be violators when the curve is near the limit.  We see that the violators are caused by normal scatter about the relation.
}

\end{figure}

\end{document}